# WEAK SINGULARITY FOR TWO-DIMENSIONAL NONLINEAR EQUATIONS OF HYDRODYNAMICS AND PROPAGATION OF SHOCK WAVES

Vitaly V. Bulatov (1), Yuriy V. Vladimirov (1), Vasily A. Vakorin (2)

(1) Institute for Problems in Mechanics Russian Academy of Sciences, Moscow, Russia.
(2) University of Saskatchewan, Saskatchewan, Canada.

**ABSTRACT:** A system of two-dimensional nonlinear equations of hydrodynamics is considered. It is shown that for the this system in the general case a solution with weak discontinuity-type singularity behaves as $\sqrt{S(x,y,t)}$, where $S(x,y,t)>0$ is a smooth function. The necessary conditions and series of corresponding differential equations are obtained for the existence of a solution.

## 0. Introduction

The role of singularities for solutions of equations of gas- and hydrodynamics is well known both in the theory of the corresponding equations and in applications. Unlike linear equations, whose solutions, at least in the small, inherit the type of singularity of the initial condition, for nonlinear equations only specific structure singularities can expand with preservation of the structure relative to perturbations of the initial conditions. V.P.Maslov appears to be first to pay attention this fact. Singularity of the type of discontinuity of the first kind of that models shock waves is the most widely known and profoundly studied. The support of this singularity is a sub-manifold of dimension one. The necessary conditions for such a solution to exist are the Hugoiniot conditions. From the present point of view these conditions are the first conditions of relations that appear in constructing the smoothness asymptotic of the solution in a neighborhood oh the discontinuity.

The whole collection of necessary conditions represents the system of ordinary differential equations on coefficients of Taylor's series relative to the distance to the singularity support. This system is not closed in the sense that the first $N$ equations contain more than $N$ unfixed parameters, and, therefore, its solutions do not lead, for example, to a unique definition of the singularity position. Although the closing of this system is possible only by taking account of global properties of the solution or under certain supplementary assumption, nevertheless, by using the numerical analysis P.Prasad discovered the following remarkable fact: if we assume the "superfluous" unfixed parameters (the last one) to be equal to zero, then the singularity support position coincides sufficient precisely with the singularity support position obtained from the precise solution of the initial system of partial differential equations[6].

There are other structural-stable singularities for solutions on nonlinear equations. In particular, there are also interesting solutions with weak discontinuity-type singularity, whose support is a sub-manifold of co-dimension greater that one (in particular, a point). V.P.Maslov established the hypothesis stating that such solutions can describe typhoon and other natural phenomena[1,2,4].

A system of two-dimensional nonlinear equations of hydrodynamics, describing the atmosphere of the Earth in the geostrophic approximation with account of the Earth's rotation is considered [1-3,5]. It is shown that for the this system in the general case a solution with weak discontinuity-type singularity behaves as $\sqrt{S(x,y,t)}$, where *S(x,y,t)>0* is a smooth function, Hess*S(x,y,t)* is not equal zero when *S(x,y,t)=0*. The necessary conditions (an analog of the Hugoiniot ones) are obtained for the existence of a solution in which such a behavior is stable under small perturbation.

In the presented paper we deduce the chain of equations which contain a number of unfixed parameters that makes it possible to obtain the link of functions which define the principal term of the asymptotic representation of the solution. As was assumed, the center of the weak singularity calculated by using the asymptotic representation for a solution of the initial nonlinear system can model the typhoon motion trajectory in the atmosphere of the rotating Earth. We performed numerical analysis of corresponding differential equations for investigating the trajectory motion of the weak point-type singularity.

If we do not consider the process of creation and destruction of vortices in the atmosphere but restrict ourselves to the description of the dynamics of such phenomena, then we can use the system of equations for the "shallow water" in which the Coriolis forces are taken into account [1-3,5]

$$\frac{\partial \eta}{\partial t} + \nabla(\eta \mathbf{u}) = 0 \qquad (0.1)$$

$$\frac{\partial \mathbf{u}}{\partial t} + (\mathbf{u}, \nabla)\mathbf{u} + \omega T\mathbf{u} + \nabla \eta = 0$$

Here $\mathbf{u} = (\mathbf{u}_1, \mathbf{u}_2)$ is the velocity vector, $\eta > 0$ is the geo-potential, $\omega/2 = \Omega$, $\Omega \sin \nu$ is the Coriolis parameter, $\Omega$ *is* the Earth's rotation frequency, *v i*s the latitude of the position, $T = \begin{pmatrix} 0 & 1 \\ -1 & 0 \end{pmatrix}$ is the matrix of rotation through $90°$, $\nabla = \left( \dfrac{\partial}{\partial x_1}, \dfrac{\partial}{\partial x_2} \right)$.

V. P. Maslov [1,2,4] conjectured that it may be possible to calculate the trajectory of a "typhoon eye" by studying the structure of singularities of solutions to quasi-linear hyperbolic equations. It was proposed to describe singular solutions in the following form (for the moment, we do not consider equations (0.1) and by *w* we denote solutions of scalar and vector equations without specifying the form of these equations)

$$w(x,t) = f(S(x,t), x, t) \tag{0.2}$$

where $f(\tau, x, t)$ is a scalar or vector function, smooth outside the point $\{\tau = 0\} \in \mathbb{R}^1$ and possessing a (possibly, weak) singularity at the point $\tau = 0$, and $S(x,t)$ *is* a smooth function. The zero set of the function $S(x,t)$ determines singular points of the solution (0.2). For example, a singularity may be a first-order discontinuity, then we have shock waves, or even be continuous and once differentiable, then we have weak discontinuities and vortex singularities.

For example, for shock waves in the one-dimensional case we have

$$w = H(x,t) + A(x,t)\Theta(x - \varphi(t)) \tag{0.3}$$

where $H(x,t)$, $A(x,t)$, and $\varphi(t)$ are smooth functions and $\Theta(t)$ is the Heaviside function. Another type of singularities is defined by the same formulas, but instead of $\Theta$ in (0.3) we have $\text{Sol}(\tau)$ equal to 0 for $\tau \neq 0$ and to 1 for $\tau = 0$. In this case *w* describes an "infinitely narrow" solution that moves on the background $H(x,t)$. From the viewpoint of generalized solutions the last solutions is equal to $H(x,t)$. However, these solution turn out to be quite reasonable if the scalar products and weak solutions are well-defined. Such solutions appear, for example, as the limiting $(as\ \varepsilon \to 0)$ solutions of the Korteweg-de-Vries equation with dispersion $\varepsilon^2$.

Finally, let us consider a two-dimensional example $(x = (x_1, x_2))$:

$$w = u^0(x,t) + u^1(x,t)(S(x,t))^\alpha \qquad (0.4)$$

where $u^0(x,t)$, $u^1(x,t)$, and $S(x,t) \geq 0$ are smooth functions, for each $t$ we have $S = 0$ only at one point $(x_1, x_2) = (X_1(t), X_2(t))$, and $\alpha > 0, \alpha \neq 0$. In this case we have a "pointwise" singularity, and the functions $X_1(t), X_2(t)$ determine the trajectory of this singularity.

Solutions of the form (0.2) (0.3) have the following important common properties.

First, they are "structurally self-similar". This means that if at a time $t_0$ these solutions have the form (0.2) with a given dependence on $\tau$, then this dependence on т is also preserved for the time $t > t_0$ at least for not very large time intervals $[t_0, t]$.

Second, these solution possess the properties of "structural stability", that is, a small variation, of the initial values for $S(x, t_0)$ or $f(\tau, x, t_0)$, which does not change the structure of the singularity of $f$ with respect to $\tau$ and the structure of the coefficients in the original equation, does not change the form and the structure of the function $w$. From the viewpoint of example (0.3), this means that, for instance, if for $t = t_0$ we replace the functions $H(x, t_0), A(x, t_0)$, and the number $\varphi(t_0)$ by some new ones, namely, by $H'(x, t_0), A'(x, t_0)$, and $\varphi'(t_0)$ then the structure of (0.3) does not change. Singularities of the form (0.4) can be considered in a similar way. It turns out that for many quasi-linear hyperbolic equations that are reasonable from the physical viewpoint, the cited structures describe almost all possible singularities with the cited properties; in this case $\alpha = 1/2$ in (0.4) [1,2,4]. This does not mean that the corresponding equation admits partial solutions that differ from the cited ones, for example, cylindrically symmetric solutions. However, such solutions rapidly disappear under the action of perturbations.

Finally, solutions of the form (0.2) have another important property, namely, they can be described by some chains of ordinary differential equations. For example, for solutions of the form (0.3), such chains are written for the coefficients $H_k$ and $A_k$ of the expansions of $H$ and $A$ into the Taylor series at the point $x = \varphi(t)$ and for the function $\varphi(t)$

$$H = \sum_{k=0}^{\infty} H_k(t)(x - \varphi(t))^k, \qquad A = \sum_{k=0}^{\infty} A_k(t)(x - \varphi(t))^k \qquad (0.5)$$

To obtain such chains, we substitute the corresponding expansions into the original equation (or a system of equations), differentiate, and set the coefficients at the summands of the equal differentiability equal to zero. Here we do not discuss how to construct the product of distributions, how to differentiate them, etc. This theory is very interesting and nontrivial [1,2,4].

For shock waves, the first equation in such a chain is the well-known Hugoiniot equation, which relates the velocity of the shock wave front and the value of the jump of the discontinuity. A general property of all such chains is that they actually form an infinite not triangular system of ordinary differential equations, that is, the first $n$ equations contain more than $n$ unknown variables. For example, for the coefficients $H_k$ and $A_k$ in the expansion (0.5) of the solution of the Hopf equation $\omega_t + \frac{1}{2}(\omega^2)_x = 0$, we have the chains

$$\dot{\varphi} = \frac{H_0^2 - (H_0 + A_0)^2}{2A_0}$$

$$\dot{A} + H_1 A_0 + A_0 A_1 / 2 = 0, \qquad \dot{H}_0 - H_1 A_0 / 2 = 0$$

(0.6)

$$\dot{A}_k + \frac{k+1}{2} \sum_{j=1}^{k+1} (2 H_j A_{k+1-j} + A_j A_{k+1-j}) = 0$$

$$\dot{H}_k + \frac{k+1}{2} \sum_{j=1}^{k+1} (H_j H_{k+1-j} - A_0 A_{k+1}) = 0$$

$k = 1, 2, \ldots$ The first relation is the Hugoiniot condition, the other relations are corrections to this condition.

Similar chains can be obtained for some other nonlinear hyperbolic equations and systems of equations, in particular, for equations of barotropic gas. It is very attractive, by using such chains, to analyse the propagation of shock waves (singularities) numerically. The first step of this procedure is to close the chain, that is, to make this chain well-defined in a way. Such problems are well-known in statistical physics. For example, they are used to derive the Boltzmann equation. A very interesting and simple method, which is even naive at first sight, was proposed in [6]. Namely, in the first $n$ equations of the chain, the variables $A_{n+1}$ and $H_{n+1}$ are set to be equal to zero. If we compare numerical results obtained for a chain closed in this way with the results of direct computations of shock waves in the original equation, we see a very good coincidence even for a small number of equations. Probably, this fact can be explained as follows. If the functions *H(x,t)* and *A(x,t)* are sufficiently smooth, then for not very large $x - \varphi(t)$ these functions are sufficiently well approximated by the Taylor series. By setting $A_{n+1}$ and $H_{n+1}$ equal to zero, we simply cut these Taylor series. Obviously, the less the time *t* is the better we can approximate the solution of the Cauchy problem by closed chains for the equations for $H(x,t), A(x,t)$ and $\varphi(x,t)$ [1,2,6].

Now let us discuss the setting of the problem for closed chains. In the problem considered we obtain a similar chain of equations that, by V. P. Maslov's conjecture, describe the motion of a "typhoon eye". Experimentally, the trajectory of this center, which is described by two functions, can be determined rather reliably, in contrast to, for example, the velocity of the air in a large body in atmosphere. We want to reconstruct the trajectory of the typhoon eye (that is, of the center of the singularity of the solution satisfying the equation of shallow water) for the time $t > t_2$ if we know its trajectory for $t \leq t_2$. In our example of a shock wave propagation, a similar problem can be stated as follows.

Suppose that we know only the function $\varphi(t), t \in [0, t_2]$, that describes the position of the shock wave front. Is it possible to find this function for $t > t_2$ if we know that $\varphi(t)$ a component of the solution, of the system (of ordinary differential equations) of the closed chain obtained from (0.6). For example, this is possible if at time $t = t_2$ we can uniquely reconstruct the values of the functions $H_0, A_0, H_1, A_1$, etc. via $\varphi(t)$ for $t \in [0, t_1]$. Then we can solve the Cauchy problem for the corresponding system (a cut chain) for the time $t > t_1$, and hence, find $\varphi(t)$. The reconstruction problem for all components of the solution of a system of ordinary differential equations via one or several components is well-known in the control theory.

We shall illustrate the above method for the chain (0.6), which are cut at the second step. We take equations with $k = 1, 2$ and set $H_2 = A_2 = 0$. The obtained chain can be easily integrated and we have

$$\tilde{\varphi} = \frac{c_3}{c_2 - c_1}\sqrt{\frac{t + c_1}{t + c_2}} + c_4 t + c_5, \text{ where } c_j, j = 1, \ldots, 5, \text{ are constants of integration.}$$

Now if we known the function $\varphi^N(t)$, describing the position of the actual shock wave front, which satisfies the Hopf equation on $t \in [t_1, t_2]$, then, by choosing the parameters $c_j$ so that on this time interval the constructed $\tilde{\varphi}$ be the best approximation of $\varphi^N(t)$ for example, in the sense of mean square error, we can extend $\varphi^N$ for the time $t > t_2$ by setting $\varphi^N \cong \tilde{\varphi}(t)$.

We point out that this example is only a simple illustration. We do not discuss whether, in the problem about the discontinuity propagation for the Hopf equation, the exact solution $\varphi(t)$ can be approximated by the function $\tilde{\varphi}$ and how the function $\varphi(t)$ changes if we increase the number of equations in the chain (0.6), by cutting it for $k = 2, 3, \ldots$ Here the results of analytic consideration are not quite clear for us. Nevertheless, for the moment, we have no answer to similar questions in the problem about the propagation of vortex singularities of two-dimensional equations in hydrodynamics.

## 1. Setting of the Problem

We shall consider solutions of equations of "shallow water" (0.1) in the form:

$$\eta = \rho(x,t) + R(x,t)F(S)$$

$$u = u(x,t) + U(x,t)F(S) \tag{1.1}$$

Here $x = (x_1, x_2) \in \mathbb{R}^2, t \geq 0, \rho, R,$ and $S$ are scalar smooth functions, $u = (v, w)$ and $U = (U_1, U_2)$ are vector smooth functions, and $S \geq 0$ and for each $t$ we have $S = 0$ only at one point $x = X(t)$. The function $F(\tau)$ (the "singularity" function) is supposed to be:

i) continuous for $\tau \geq 0$,

ii) smooth and strictly positive for $\tau > 0$; $F(0) = 0$ and $F'(\tau) \to \infty$ as $\tau \to +0$. The set $\Gamma = (x = X(t), t \in [0, T])$ is called the *trajectory* of the singular solution (1.1) satisfying equation (0.1) on the interval $[0, T]$.

iii) the matrix of second-order derivatives $\left\| \dfrac{\partial^2 S}{\partial x_i \partial x_j} \right\|_\Gamma = \text{Hess } S \big|_\Gamma$ is required to be non-degenerate and, hence, positive on $\Gamma$.

Finally, the function $S$ is required to satisfy the condition. of "general position" so that iv) the form $(y, \mathrm{Hess}\, S|_\Gamma y)$ be not proportional to $y_1^2 + y_2^2$. (In what follows, we show that if solution (1.1) exists and the cited conditions on $S(x,t)$ are satisfied at time $t = t_0$ then these conditions are satisfied at any subsequent time).

We consider the problem of constructing solutions (1.1) or, at least, some characteristics of these solutions: first of all, the functions $F(\tau)$, the trajectories $\Gamma$, the values of the functions $\rho, R, u,$ and $U$ on $\Gamma$, some of their derivatives, $\mathrm{Hess}\, S,$, etc., that is, some characteristics of the solution in a neighborhood of the trajectory $\Gamma$. Though we deal with some special solutions of system (0.1), there exist many such solutions and, in fact, we construct a family, depending on some parameters; we denote them by $\gamma = (\gamma_1, \gamma_2 \ldots)$. Generally speaking, there may be infinitely many such parameters.

Suppose that we have constructed a class of such solutions and that the trajectory of singularities is described by $X(t, \gamma) = (X_1(t, \gamma), X_2(t, \gamma))$. Then, by taking into account V. P. Maslov's conjecture that the solutions (1.1) describe mesoscaled vortices and knowing the trajectory of the vortices for the time $[0, T]$ we can reconstruct the trajectory of the future motion of the vortices as follows. Let $\Gamma_N = (x_1 = X_1^N(t), x_2 = X_2^N(t), t \in [t_1, t_2])$ be the trajectory of an actual vortices. We choose the parameters $\gamma = (\gamma_1, \gamma_2 \ldots)$ from the conditions that the trajectories $\Gamma$ and $\Gamma_N$ are close to each other on the time interval $[0, T_1]$, for example, from the minimization condition for the mean square error

$$\int_0^{T_1} (X_1^N(t) - X_1(t, \gamma))^2 + (X_2^N(t) - X_2(t, \gamma))^2 \, dt \to \min. \qquad (1.2)$$

If we know the parameters $\gamma$, we can uniquely determine the trajectory $\Gamma$ also for the time $t > t_2$ The number of parameters $\gamma$ may be sufficiently large, and we must choose the number of them from the following considerations: on the one hand, the class of functions $X(t, \gamma)$ must approximate the trajectory of the vortex sufficiently well, therefore, as we shall see in what follows, the more parameters $\gamma$ we have the better. On the other hand, an increase of the number of parameters $\gamma$ implies considerable analytic and calculating difficulties, so it is natural to try to choose a reasonable number of parameters. We choose them by using "asymptotic" considerations, that is, so that the obtained approximate solution of the initial system were, in some sense, the leading term of the formal asymptotic solution.

## 2. Hugoiniot-type Solutions Necessary for the Existence of Vortex Solutions

Now let us state the results. First of all, we present a statement that describes the non-smooth part of the solution (1.1), which was stated earlier (in a somewhat different form) and proved by V. P. Maslov for $\omega = 0$ [1,2,4].

**Proposition** I *If system* (0.1) *has a solution of the form* (1.1), *satisfying conditions* i)-iv), *then*

**1a.** $F = \sqrt{\tau}$, *so that*

$$\eta = \rho(x,t) + R(x,t)\sqrt{S(x,t)}, \qquad u = u(x,t) + U(x,t)\sqrt{S(x,t)}. \qquad (2.2')$$

**1b.** *The trajectory* $X(t)$ *is "frozen" into the field of velocities* u:

$$\dot{X}(t) = u(X(t),t) \qquad (2.3)$$

**1c.** *On the trajectory* $X(t)$ *the complex velocities* $v(x,t) + iw(x,t)$ *(and* $u_1(x,t) + iu_2(x,t)$*) satisfy the Cauchy-Riemann conditions*

$$\frac{\partial v}{\partial x_1} = \frac{\partial w}{\partial x_2}, \qquad \frac{\partial v}{\partial x_2} = \frac{\partial w}{\partial x_1} \qquad \text{for } x=X(t) \qquad (2.4)$$

**1d.** *For functions* $\tilde{\rho} = R\sqrt{S}$ *and* $\tilde{u} = U\sqrt{S}$ *we have*

$$\begin{pmatrix} \tilde{u} \\ \tilde{\rho} \end{pmatrix} = A(t)\sqrt{Q^*(t)(x - X(t)), B_0 Q^*(t)(x - X(t)) + O(|x - X(t)|^3)} \times$$

$$\times \left( \begin{pmatrix} Q(t)T\, B_0\, Q^*(x - X(t)) \\ 0 \end{pmatrix} + O(|x - X(t)|^2) \right). \qquad (2.5)$$

*Here* $Q^* = \begin{pmatrix} \cos\Theta & -\sin\Theta \\ \sin\Theta & \cos\Theta \end{pmatrix}$ *is the matrix of rotation through the angle*

$$\Theta(t) = \Theta_0 + \int_0^t p(t)\,dt \equiv \Theta_0 - \frac{1}{2}\int_0^t \operatorname{rot} u(X(t),t)\,dt, \qquad (2.6)$$

$$A(t) = \exp\left(-\frac{3}{2}\int_0^t \operatorname{div} u(X(t),t)\,dt\right) \equiv$$

$$\equiv \exp\left(-3\int_0^t q(t)\,dt\right) \equiv \left(\frac{\rho(X(t),t)}{\rho(X(0),0)}\right)^{3/2} \qquad (2.7)$$

$B_0 = \begin{pmatrix} b_1 & 0 \\ 0 & b_2 \end{pmatrix}$, where $b_1 > 0, b_2 > 0, b_1 \neq b_2$, and $\Theta_0$ are real constants

*characterizing the structure of the vortex solution. We write*

$$q(t) = \frac{\partial v}{\partial x_1}(X(t),t) \equiv \frac{\partial w}{\partial x_2}(X(t),t) \equiv \frac{1}{2}\operatorname{div} u(X(t),t), \qquad (2\text{-}8)$$

$$p(t) = \frac{\partial v}{\partial x_2}(X(t),t) \equiv -\frac{\partial w}{\partial x_1}(X(t),t) \equiv \operatorname{rot}\frac{1}{2}u(X(t),t), \qquad (2.9)$$

**1e.** *For the derivatives* $\rho_{lj}^{(l+j)} = \frac{1}{l!j!}\frac{\partial^{l+j}v}{\partial x^l \partial x^j}(X(t),t),$

$$v_{lj}^{(l+j)} = \frac{1}{l!j!}\frac{\partial^{l+j}v}{\partial x_1^l \partial x_2^j}(X(t),t), \qquad w_{lj}^{(l+j)} = \frac{1}{l!j!}\frac{\partial^{l+j}w}{\partial x_1^l \partial x_2^j}(X(t),t), \qquad (2.10)$$

*and* $v_0^{(0)} = V_1, w_0^{(0)} = V_2$, *in addition to (2.3) and (2.4), we have the conditions (the initial conditions from the "Hugoiniot chain")'.*

$$\dot{X}_1 = V_1, \quad \dot{X}_2 = V_2, \qquad (2.11)$$

$$\dot{V}_1 + \omega V_2 + \rho_{10}^{(1)} = 0, \qquad \dot{V}_2 + \omega V_1 + \rho_{01}^{(1)} = 0, \qquad (2.12)$$

$$\dot{\rho}_0^{(0)} + 2q\rho_0^{(0)} = 0, \qquad (2.13)$$

$$\begin{cases} \dot{\rho}_{10}^{(1)} + 3q\rho_{10}^{(1)} - p\rho_0^{(1)} + \rho_0^{(0)}(w_{11}^{(2)} + 2v_{20}^{(2)}) = 0, \\ \dot{\rho}_{01}^{(1)} + 3q\rho_{01}^{(1)} - p\rho_{10}^{(1)} + \rho_0^{(0)}(v_{11}^{(2)} + 2w_{02}^{(2)}) = 0, \end{cases} \qquad (2.14)$$

$$\dot{q} - p^2 + q^2 + \omega p + 2r = 0, \qquad (2.15)$$

$$\dot{p} + 2pq - \omega q = 0, \qquad (2.16)$$

$$\dot{r} + 4pr + \frac{1}{2}\rho_{10}^{(1)}(3v_{20}^{(2)} + w_{11}^{(2)} + v_{02}^{(2)}) + \frac{1}{2}\rho_{01}^{(1)}(v_{11}^{(2)} + 3w_{02}^{(2)} + w_{20}^{(2)}) =$$

$$= -\rho_0^{(0)}(3v_{30}^{(3)} + 3w_{03}^{(3)} + w_{21}^{(3)} + v_{12}^{(3)}), \qquad (2.17)$$

$$\begin{cases}
\dot{v}_{20}^{(2)} + 3qv_{20}^{(2)} - \omega\, w_{20}^{(2)} - p(v_{11}^{(2)} - w_{20}^{(2)}) &= -3\rho_{30}^{(3)} \\
\dot{v}_{11}^{(2)} + 3qv_{11}^{(2)} - \omega\, w_{11}^{(3)} - p(2v_{02}^{(2)} - 2v_{20}^{(2)} - w_{11}^{(2)}) &= -2\rho_{21}^{(3)} \\
\dot{v}_{02}^{(2)} + 3qv_{02}^{(2)} - \omega\, w_{02}^{(2)} + p(v_{11}^{(2)} + w_{02}^{(2)}) &= -\rho_{12}^{(3)} \\
\dot{w}_{20}^{(2)} + 3qw_{20}^{(2)} + \omega\, v_{20}^{(2)} - p(w_{11}^{(2)} - v_{20}^{(2)}) &= -\rho_{21}^{(3)} \\
\dot{w}_{11}^{(2)} + 3qw_{11}^{(2)} + \omega\, v_{11}^{(2)} - p(-2w_{20}^{(2)} + 2w_{02}^{(2)} + v_{11}^{(2)}) &= -2\rho_{12}^{(3)} \\
\dot{w}_{02}^{(2)} + 3qw_{02}^{(2)} + \omega\, v_{02}^{(2)} + p(w_{11}^{(2)} - v_{02}^{(2)}) &= -3\rho_{03}^{(3)},
\end{cases} \qquad (2.18)$$

$$\rho_0^{(0)}(3v_{30}^{(3)} + w_{21}^{(3)} - v_{12}^{(3)} - 3w_{03}^{(3)}) + \rho_{10}^{(1)}(3v_{20}^{(2)} + w_{11}^{(2)} - v_{02}^{(2)}) -$$

$$- \rho_{01}^{(0)}(v_{11}^{(2)} + 3w_{02}^{(2)} - w_{20}^{(2)}) = 0 \qquad (2.19)$$

**2.** *Conditions* (2.3), (2.4), (2.11)-(2.18), *and representation* (2.5)-(2.7)

*is necessary and sufficient for the functions* (1.1) *to satisfy the initial equations* (0.1) *with accuracy*

$O(|x - X(t)|^3)$.

The relations presented in Proposition 1 are some analogs of Hugoiniot conditions and corrections to them for solitary vortices in equations of "shallow water" (0.1).

Here we have no room to present the proof of this statement. We only note that condition (2.3), which means that the vortex is "frozen", and the Cauchy-Riemann condition (2.4) follow from the fact that there is a non-smooth part in solution (1.1). These conditions can be obtained similarly to the ray method or the WKB method applied to the expansion with respect to smoothness. Note that to derive these conditions we essentially use statement iv) about the "anti-symmetry" of the vortex. In what follows, by general considerations, we obtain relations similar to (2.5)-(2.7) from higher-order Taylor expansions for $S, U$, and $R$. To obtain conditions (2.ll)-(2.19), we successively differentiate the initial system with regard for (2.3)-(2.4) and calculate the result at points of the trajectory $x = X(t)$. In fact, this part of the proof only requires very cumbersome calculations.

Formulas (2.5)-(2.7) are transparent and can be easily analyzed. The function $\widetilde{u}$ describes the motion of a solitary vortex in the field of velocities $u(x,t)$. This fact is well-known in hydrodynamics. So, since this fact takes place for solutions of the form (1.1), this solution does not contradict the laws of hydrodynamics. Nevertheless, this fact must be proved for solutions of the form (1.1). It is rather curious that the Cauchy-Riemann conditions (2.4) must be satisfied. This condition is not invariant for (all) trajectories of the field of velocities, and thus, shows how a vortex effects the "smooth background" $u(x,t)$. The vortex itself moves along the trajectory $X(t)$ and rotates (due to the Cauchy-Riemann conditions) with angular velocity $\dot{\theta} = \frac{1}{2}\mathrm{rot}\, u(X(t),t)$. In the main, the "section" of the vortex is an ellipse with half-axes given by $b_1$ and $b_2$ and the initial angle $\Theta_0$. We note that with the growth of the distance from $X(t)$ the "vortex" (non-smooth) part of the solution increases rather slowly, that is, as $|x - X(t)|^2$, and the function (2.5) increases as $|x - X(t)|^3$. This slow increase of (2.5) is an argument to use these functions for the description of typhoons, since it shows that the typhoon has an "eye". Finally, note that the described dynamics of the vortex remains asymmetric. In the first approximation, the vortex is conformally transformed: it equally rotates, contracts, and extends in all directions, the extension coefficient is determined by the divergence of the field of velocities $u(x,t)$ on the trajectory $X(t)$.

## 3. Closing of the Chain

We see from relations (2.17)-(2.18) that along with (2.11)-(2.16) they form an underdetermined system of differential equations. If we write necessary conditions that relate the next Taylor coefficients for the functions $u, \rho, S, U,$ and $R,$ then this situation preserves. In other words, we obtain a non-closed chain of equations. We have a problem how to close this system. As we have already mentioned, the problem of closing chains is one of the most complicated and interesting problems in mathematical physics. Recall that, in particular, by closing the chains describing the density distribution functions, we obtain the Boltszmann equation. Here we close the chain as in the papers [11, 22]. For the function $\rho$ we set the summands with coefficients at the third-order polynomials equal to zero and assume that

$$3v_{30}^{(3)} + v_{12}^{(3)} + 3w_{03}^{(3)} + w_{21}^{(3)} = 0 \tag{3.20}$$

and

$$\rho_{30}^{(3)} = \rho_{21}^{(3)} = \rho_{12}^{(3)} = \rho_{03}^{(3)} = 0. \tag{3.21}$$

In other words, we set the functions on the right-hand sides in (2.17)-(2.18) equal to zero. (In conjunction with (2.19), relation (2.20) means that $v_{30}^{(3)}$, $w_{21}^{(3)}, w_{03}^{(3)}$, and $v_{12}^{(3)}$ are related to the preceding coefficients:

$$3w_{30}^{(3)} + w_{21}^{(3)} = -(3w_{03}^{(3)} + v_{12}^{(3)}) =$$

$$= \frac{1}{2}\rho_{10}^{(1)}(3v_{20}^{(2)} + w_{11}^{(2)} - v_{02}^{(2)}) - \frac{1}{2}\rho_{01}^{(1)}(3w_{02}^{(2)} + v_{11}^{(2)} - w_{20}^{(2)}) \qquad (3.22)$$

generally speaking, this means that it is impossible to set all third-order derivatives of the functions *v, w,* and *p* equal to zero and to preserve the discrepancy $O(|x - X(t)|^3)$ in (0.1)). Now we obtain a closed system of ordinary differential equations.

**Proposition 2** *Suppose that the coefficients* $\rho_k^{(3)}, u_k^{(3)}, v_k^{(3)}, k = (k_1, k_2), |k| \equiv k_1 + k_2 = 3$, *are chosen by the above method, the coefficients* $\rho_k^{(|k|)}$, $v_k^{(|k|)}$, *and* $w_k^{(|k|)}$ *satisfy equations* (2.11)-(2.18) *for* $|k| \leq 2$, *and* $S, U, R$ *have the form* (2.5)-(2.7). *Then the functions* (1.1) *satisfy the original system with accuracy* $O(|x - X(t)|^3)$ *for all values of the Taylor coefficients for the functions* $u = (v, w), U = (U_1, U_2), \rho, R$, *and* $S, k = (k_1, k_2); |k| \geq 4$ *for* $v, w$, *and* $\rho, |k| \geq 2$ *for U and* $\rho |k| \geq 3$ *for S. In particular, by setting all "higher-order" coefficients equal to zero, we obtain an approximate solution of the form* (1.1) $\mod O(|x - X(t)|^3)$ *with* $u, \rho, U, R,$, *and S of polynomial form (with powers less than or equal to* 3*) depending on* $(x - X(t))$.

Thus, after the chain is cut, we obtain, in a sense, an asymptotic solution (with respect to powers of $x - X(t)$) of the initial equation.

We do not know, at least for the moment, how to prove rigorously the fact that, by closing the chain as proposed above, we obtain the trajectory of the singularity *X(t)* that is close to the actual trajectory of the vortex for some times. However, as we shall see in what follows, the desired solution. of the closed chain must be stable (we choose only stable solutions). Therefore, if we assume that the right-hand sides in (2.17) and (2.18) do not vanish but are sufficiently small, then, probably, one can prove that their contribution to the solution is small. On the other hand, it is required to have at least the discrepancy $O(|x - X(t)|^3)$ on the right-hand side of (0.1) to determine the "higher-order" part of the non-smooth component of solution (1.1). Therefore, we have cut the

chain at a reasonable step; if we cut it at the previous step (that is, we set the coefficients $v_{20}^{(2)}$, $w_{20}^{(2)}$, $v_{02}^{(2)}$, etc. equal to zero), then it is impossible to give a true description of the higher-order non-smooth part of the solution. If we consider more equations, then we obtain much more complicated equations for the Taylor coefficients. Numerically and analytically, we can show that if we consider only the first- and second-order summands, we obtain a different behavior of the system that contains second-order summands.

## 4. Reduction of a "Cut-off" Chain to the Hill Equation

It is remarkable and rather unexpected that the system (2.11)-(2.21) can be reduced to the Hill equation

$$\frac{d^2 y}{d\Phi^2} + q(\Phi, \beta, \Omega_0) y = 0. \qquad (4.23)$$

Here

$$q = \Omega_0^2 + \mathrm{Re}\left(\frac{\beta_0 \beta_1}{2} \exp(2i\Phi) + \overline{\beta_0 \beta_1} \exp(-2i\Phi) + \beta_1 \beta_2 \exp(i\Phi) - \overline{\beta_0} \beta_2 \exp(-i\Phi)\right),$$

complex parameters $\beta_0, \beta_1$, and $\beta_2$, and a real parameter $\Omega_0 \geq 0$ are constants of integration, $\beta = (\beta_0, \beta_1, \beta_2)$.

According to the general theory of the Hill equation, the solutions of equation (4.23) may behave as follows:

*Stable case.* We can compose the basis of solutions of equation (4.23) from solutions of the form $y = y_0 \exp(i\Omega\Phi), \overline{y} = \overline{y}_0 \exp(-i\Omega\Phi)$, where $\Omega$ is a real number called a *characteristic index* and $y_0(\Phi, \beta, \Omega_0)$ is a complex smooth function, which is $2\pi$-periodic with respect to $\Phi$ and does not vanish. The functions $y_0$ and the number $\Omega$ are determined up to a normed coefficient and up to a sign of $\Omega$. Moreover, to $\Omega$ one can add any number multiple of $\pi$. We shall fix our choice of $y_0$ and $\Omega$ as follows.

We shall need

$$\overline{y} \frac{\partial y}{\partial \Phi} - y \frac{\partial \overline{y}}{\partial \Phi} = 2i, \qquad (4.24)$$

that is, we need that the solution $y$ corresponds to second-order multiplicators [1,2,4]. In particular, it immediately implies that $y \neq 0$ and that the smooth functions $g(\Phi) = |y|$ and $\theta = \text{Arg}\, y(\Phi)$ are defined; in this case they satisfy the relation $\dfrac{d\theta}{d\Phi} = 1/g^2(\Phi)$. Moreover, we can take $\Omega$ in the form

$\Omega = \dfrac{1}{2\pi}\int_0^{2\pi} d\Phi / g^2(\Phi)$ The argument $\theta$ can be completely determined from the relation $\Theta|_{\Phi=0} = 0$ thus, we obtain

$$\theta = \int_0^{\Phi} d\Phi / g^2(\Phi) \qquad (4.25)$$

Obviously, $\theta = \Omega\Phi + \theta_0(\Phi)$ where $\theta_0$ is a $2\pi$-periodic function of $\Phi$. The number $\Omega$ is called a *quasi-momentum*. Equation (4.23) is called *strongly stable* if $\Omega \neq n/2$, $n = 0, 1, 2$; in this case, the property of being stable is preserved for small variations of the parameters $\beta$ and $\Omega$. If $\Omega = n/2$, then the solutions $y$ and $\bar{y}$ turn out to be periodic or anti-periodic functions'. Nevertheless, the stability property is not preserved if the parameters $\beta$ and $\Omega$ change.

*Non-stable* case. In this case, at least one of the solutions of equation (4.23) increases infinitely (linearly or nonlinearly) as $\Phi \to \infty$. Now let us state the main result about the cut chain (2.11)-(2.21).

**Proposition 3** (Reduction of the problem about the pointwise vortex singularity to the Hill equation). *Let $\omega > 0$, then:*

**1a.** *If the parameters $\beta$ and $\Omega_0$ lie in the region of "exponential" non-stability, then there exists a (infinite) sequence of time instants $t_1, t_2, \ldots, t_m, \ldots$, such that the sequence $\rho_0^{(0)}|_{t=t_m}$ either grows exponentially fast or decreases exponentially fast.*

**1b.** *If the parameters $\beta$ and $\Omega_0$ lie in the stability region, then for any t the component $\rho_0^{(0)}(t)$ is strictly bounded below and above by positive constants. In this case, $\rho_0^{(0)}$, the velocity components $V_1$ and $V_2$, and the trajectory $(X_1, X_2)$ are expressed in terms of the functions $g(\Phi)$ and $\theta(\Phi)$, that is, via the solutions of the Hill equation* (4.23) *as follows:*

$$\rho_0^{(0)} = \frac{\omega \mu g^2(\Phi)}{2|c|(1-(1-\mu^2)\sin^2(\omega(t-t_0)/2))} \tag{4.26}$$

$$V_1 = \mathrm{Re}\, V, \quad V_2 = \mathrm{Im}\, V \quad V = \exp(-i\omega t)(\frac{i}{4}\sqrt{\frac{\omega}{\mu|c|}} I_+ + V^0) \tag{4.27}$$

$$X_1 = \mathrm{Re}\, X_1, \qquad X_2 = \mathrm{Re}\, X;$$

$$X = X^0 - \frac{1}{4}\sqrt{\frac{1}{\mu|c|\omega}} (\exp(-i\omega t) I_+ + I_-) + \frac{i}{\omega}(\exp(-i\omega t) - 1) V^0$$

Here $V^0 = V_1^0 + i V_2^0$ and $X^0 = X_1^0 + i X_2^0$ are complex constants of integration, $t_0, c,$ and $\mu$ $(c \neq 0, 0 < \mu \leq 1)$ are real constants of integration,

$$I_{\pm} = \exp(\pm\frac{i\omega t_0}{2}) \int_0^{\Phi(t)} g(\Phi)(\mu \cos(\theta(\Phi) - \Phi_0) \pm i \sin(\theta(\Phi) - \theta_0))$$

$$\times \exp(-\frac{i\Phi}{2})(\beta_0 \exp(i\Phi) - 2\beta_1 \exp(-i\Phi) + 2\beta_2) d\Phi, \tag{4.29}$$

and $\Phi(t)$ related to $t$ by the formula

$$\Theta(\Phi) = \mathrm{Arctg}\left(\mu\, \mathrm{tg}\left(\frac{\omega}{2}(t-t_0)\right)\right) + \Theta_0, \qquad \Theta_0 = \mathrm{Arctg}\,\mu\, \mathrm{tg}\left(\frac{\omega}{2} t_0\right) \tag{4.30}$$

By Arctg we *denote the value for which, in the first case,* $\theta(\Phi) - \theta_0$ *lies in the same interval* $[\pi n, \pi(n+1)]$ *as* $\omega/2(t-t_0)$ *and, in the second case, во lies in the same interval* $[\pi n, \pi(n+1)]$ *as* $\omega t_0/2$

2. *Let* $\omega = 0$, *then* $\rho_0^{(0)} \sim 1/t^2$ *for* $t \gg 1$.

So, formulas (4.26)-(4.28) define a class of functions that depend on five complex and four real parameters. We shall look for the trajectory of vortex singularities of equation (0.1). We also note that the functions $g\cos(\theta - \Phi_0)$ and $g\sin(\theta - \theta_0)$ satisfy the Hill equation (4.23). Hence, we can also say that the above formulas define a (nonlinear) transformation that connects the cut chain (2.11)-(2.21) with the linear equation (4.23). Obviously, the function $\rho_0^{(0)}$ is a almost periodic function with two periods; this function depends on three complex parameters $\beta_0, \beta_1,$ and $\beta_2$ and on four real parameters $\Omega_0 > 0$, $c \neq 0$, $0 < \mu \leq 1$ and $t_0 \in [0, 2\pi]$. The functions $V_1$ and $V_2$ depend on an additional complex parameter $V^0$, and the functions $X_1$ and $X_2$ depend also on the parameter $X^0$.

Finally, we present here the formula that relates rot $u$ on the trajectory and $\rho_0^{(0)}$:

$$\text{rot } u|_{x=X(t)} = -c\rho_0^{(0)} - \omega \qquad (4.31)$$

By its physical meaning, $\rho_0^{(0)}$ must be a positive bounded function that for all $t$ is bounded away from 0 by a positive number. Therefore, if the Coriolis force is absent, the chain (2.11)-(2.21) has no physically interesting solutions. Thus, physically reasonable stable solutions $\rho_0^{(0)}$ are possible because of the presence of the Coriolis forces. Note that near the equator, where $\omega = 0$ typhoons have never been observed [3,5].

*First approximation.* Generally speaking, formulas (4.28)-(4.30) define a sufficiently fast oscillating function if time interval, in which we are interested in., contains some intervals $(0, 4\pi/\omega)$. The amplitude of these oscillations is the greater the more is the difference between the potential $q$ and a constant. Since the experimental trajectory, which we want to approximate by a function of the form (4.28)-(4.30), is described by a smooth curve, it is natural to assume that the parameters $\beta_0, \beta_1,$ and $\beta_2$ in (4.28)-(4.29) are sufficiently small and integrate the Hill equation by the averaging methods. Then if we write $|\beta| = \sqrt{|\beta_0|^2 + |\beta_1|^2 + |\beta_2|^2}$, in the first approximation we obtain

$$g = \frac{1}{\sqrt{\Omega_0}} + O(|\beta|^2), \qquad \theta = \Omega_0 \Phi + O(|\beta|^2)$$

We substitute $g$ and $\theta$ into (4.28)-(4.30), perform the calculations, combine some terms that appear after the integrals $J^{\pm}$ with $X^0$ and $V^0$ are calculated, and arrive at the following statement.

**Proposition 4** *Suppose that $\omega > 0$, $\beta$ and $\Omega_0$ belong to the stability region of equation (4.23) and the $|\beta|$ is sufficiently small, then the function X has the form*

$$X = A_1 + A_2 \exp(-i\omega t) + \frac{4ic\sqrt{\Omega_0}}{\omega} \exp(-i\omega t/2) \sqrt{(\mu^2 - 1)\sin^2 \frac{\omega(t - t_0)}{2} + 1}$$

$$\times \exp(i\Phi/2) \left( \frac{\beta_0}{\Omega_0^2 - 1/4} - \frac{2\overline{\beta_1} \exp(-2i\Phi)}{\Omega_0^2 - 9/4} - \frac{\beta_2 \exp(-i\Phi)}{\Omega_0^2 - 1/4} \right)$$

$$+ O\left( \frac{|\beta|^3}{\min(\Omega_0^2 - 1/4, \Omega_0^2 - 9/4)} \right). \qquad (4.32)$$

*Here $A_1$ and $A_2$ are "new" complex constants of integration, and $\Phi = \Phi(t)$ can be calculated by the formula*

$$\Phi = \frac{1}{\Omega_0} \text{Arctg}\left( \mu \, \text{tg} \frac{\omega(t - t_0)}{2} \right) + \frac{1}{\Omega_0} \text{Arctg}\left( \mu \, \text{tg} \frac{\omega t_0}{2} \right). \qquad (4.33)$$

Note that in the last formula for *X* the number of constants is reduced by one, since the parameter c can be included into $\beta_0, \beta_1$, and $\beta_2$. This formula shows that for small $|\beta|$ all summands in (4.32), except $A_1$ and the summand that contains $\exp(\frac{i(\Phi - \omega t)}{2})\beta_0/(\Omega_0^2 - 1/4)$, , oscillate with frequency greater than *w/2*. By definition of the function $\Phi(t)$, we easily obtain $\Phi = \frac{\omega t}{2\Omega_0} + \widetilde{\Phi}(t, \mu, t_0)$, where $\widetilde{\Phi}$ is a periodic function with period $T = \frac{\pi}{\omega}$; therefore the whole function *X* (4.32) oscillates sufficiently fast if the frequency $\Omega_0$ differs strongly from 1/2, and such a function badly approximates the trajectory of the "typhoon eye". Thus we conclude: the frequency $\Omega_0$ must be close to 1/2.

Now, instead of the free parameters $\beta_0$, $\beta_1$, and $\beta_2$, we introduce the new parameters

$$A_0 = \frac{4ic\sqrt{\Omega_0}\,\beta_0}{\Omega_0^2 - 1/4}, \qquad A_3 = -\frac{8i\overline{\beta_1}c\sqrt{\Omega_0}}{\omega(\Omega_0^2 - 9/4)}, \qquad A_4 = -\frac{4ic\sqrt{\Omega_0}\,\beta_2}{\omega(\Omega_0^2 - 1/4)}.$$

Then the complex vector-function that determines the trajectory takes the form:

$$X \equiv X(t, \Omega_0, \mu, t_0, A) \approx A_1 + A_2 \exp(-i\omega t) +$$

$$+ \exp(i\frac{(\Phi - \omega t)}{2})(A_0 + A_3 \exp(\exp(-2i\Phi)) + A_4 \exp(-i\Phi))\sqrt{1 - (A - \mu^2)\sin^2 \frac{\omega(t - t_0)}{2}} \quad (4.34)$$

where $\Phi = \Phi(t, \mu, t_0, \Omega_0)$ is determined by (4.33). The above arguments imply that this function can approximate a smooth trajectory if $\Omega_0$ is close to 1/2 and the module of complex parameters $A_2, A_3$, and $A_4$ turn out to be less than the module of the parameters $A_0$ and $A_1$. Note that this requirement does not contradict the assumption that $\beta_0, \beta_1$ and $\beta_2$ are small. This means that $|\beta_0| \sim |\Omega_0^2 - 1/4|$, and that $|\beta_2|$ has the order of smallness less than that of $|\beta_0|$. If we reconstruct the trajectory of an. actual typhoon, we determine the parameters $A, \Omega_0, \mu,$ and $t_0$ by condition (2.2).

In conclusion, taking into account the above considerations, we perform one more approximation. For this purpose, in (4.34) we omit the summand that includes the coefficients $A_2, A_3,$ and $A_4$. Moreover, if we assume that $1 - \mu$ is not very large, then we obtain $X \approx A_1 + \exp(i(\Phi - \omega t)/2)A_0$. On the plane $(X_1, X_2)$ this equation describes a circle with center at the point $(ReA_1, ReA_2)$ and radius $|A_0|$. If the angular velocity

$$\dot{\Phi} - \omega = \omega \left( \frac{\mu}{2\Omega_0(1 - (1 - \mu^2)\sin^2(\omega(t - t_0)/2))} - 1 \right)$$

is nonnegative, then the motion is counter-clockwise, if non-positive, then clockwise. The angular velocity is nonnegative (the motion is counterclockwise) if, at least, $\mu > 2\Omega_0$ and non-positive (the motion is clockwise) if $\mu < 2\Omega_0$. Note that, since we suppose that $\Omega_0$ is close to 1/2, the velocity of the motion along the circle is essentially less than $\omega$, and during some periods $T = 2\pi/\omega$ the "vortex singularity" goes only through a part of the circle.

The results obtained make it possible to established the following hypothesis, which may be basis for a prediction method for the expansion of a real mesoscaled vortices. Let us assume that there exists a system of equation that model the singularities dynamics and let its solutions (similar to those we have considered in the present paper) correspond to a real vortices. If we know the real vortices trajectory on a some initial time interval and choose the initial conditions for the corresponding differential equations in such a way that the real and calculated trajectories are close on this initial time interval, then we can suppose that the calculated vortices trajectory also is close to the real one on the next time interval.

This work supported by Russian Foundation of Basic Research, Grant 99-01-00856.